\documentclass[12pt]{spieman}  
\usepackage{amsmath,amsfonts,amssymb}
\usepackage{graphicx}
\usepackage{setspace}
\usepackage{tocloft}
\usepackage{caption}
\usepackage{gensymb}
\usepackage{xcolor}
\usepackage{lineno}

\title{Equilibrium positions of optically-trapped single-crystal uniaxial calcite}

\author[a]{Elaina M Wahmann}
\author[a*]{Catherine M Herne}
\affil[a]{SUNY New Paltz, Department of Physics and Astronomy, 1 Hawk Drive, New Paltz, NY 12561}

\cftpagenumbersoff{figure}
\cftpagenumbersoff{table} 

\begin{document} 
\maketitle

\begin{abstract}

Precision in optical micromanipulation is critical for non-invasive, non-contact control of objects with light and for using light to measure forces and torques. We demonstrate the control of birefringent objects over three translational directions and three rotational degrees of freedom. Our work trapping and rotating calcite, a unique crystal with no axis of symmetry, extends the positioning abilities of optical tweezers. We trap and levitate a single crystal in linearly polarized light and position it in three rotational directions relative to the polarization. Our model of the torque on the crystals is based on determining the induced polarization and its interaction with the incident electric field. Crystals are shown to rest in equilibrium with their optic axis along the beam axis and one extraordinary axis along the direction of polarization. The calculated equilibrium positions align with our observations of stable trapped calcite. 

\end{abstract}



\section{Introduction}

The field of optical tweezers has opened avenues to trapping and manipulating a multitude of nano- and micro-scale objects. These have been held and moved due to light’s scattering and gradient forces, and spun due to light polarization and optical vortices. The ability to manipulate small objects in three dimensions has been applied to biological systems; for example, measuring forces between cells \cite{Dienerowitz2014}, the index of refraction of cells and viruses \cite{Nagesh2014,Pang2016}, and properties of biomolecules and molecular motors \cite{Spudich2011}. The “lab-on-a-chip” concept uses optical tweezers to conduct several operations in a small space, such as a microscope slide \cite{Mushfique2008}. Microscopic sorters use optically rotated “fans” to sort cells by size \cite{Macdonald2003}. These are only a tiny fraction of the methods and applications of optical tweezers. The advantages of optical tweezers over other means of micromanipulation are their small size, ability to work in fluids, and non-destructive properties. This work demonstrates a novel improvement in manipulation. We show experimentally the means to precisely position objects of interest in three dimensions and in three rotational degrees of freedom.

Objects are typically rotated by applying either a circularly polarized beam to birefringent objects or an optical vortex to reflecting or absorbing objects. The response of birefringence to polarized light is well-characterized. Beginning with shards of birefringent crystals \cite{Friese1998} and moving to spherical crystals \cite{Parkin2009} and liquid crystals \cite{SYDG2013}, there have been extensive studies of alignment with linearly polarized light and rotation with circularly polarized light. Several results show that birefringent crystals will orient with their fast axis along the electric field \cite{WCC2008, Stilgoe2022}. The shape-birefringence of structures that have asymmetry in shape but not in their crystal structure has also been examined both theoretically and experimentally. Cylindrical shapes were shown to orient themselves with their longest axis along the beam \cite{Gauthier1997, Gauthier1999, Cao2012}, and as the aspect ratio of the width to length increased, they maintained the longest diagonal along the beam axis. Cubic shapes have one stable equilibrium with their corners aligned with the beam axis (also along the longest diagonal) \cite{Gauthier2000}. Combining shape and internal birefringence, microstructured objects have been controlled and rotated with linear and circular polarization, incorporating the interplay of torques due to the shape and internal birefringence \cite{Higurashi1998, Higurashi1999, Singer2006}. 

These investigations using polarized light interacting with birefringent objects have been limited to birefringent crystal fragments, birefringent spheres, and micro-fabricated birefringent crystals. These either have no defined shape, are spherical, or have axial symmetry. Our rhombohedral calcite crystals, however, are near-cubic, with their longest dimension between the two acute corners. They have internal birefringence along three different principal sections. The optic axis does not run along a physical line of symmetry. All of these factors mean that while the shape is clearly defined, it is not a symmetric structure. Hence, we can determine their position with respect to all six degrees of freedom. When attached to a biological cell or filament, then, these crystals could act as “handles” for precise positioning. For example, an attached cell could be rotated to be oriented along the axis of a fixed cell, simply by rotating the polarization direction, in order to measure cell-cell interaction forces. Controlling objects over three translational and three rotational degrees of freedom would bring more micromachines like those discussed above into reality. 

Previous studies of the dynamics of the rotational motion of calcite crystals due to spin angular momentum have been performed \cite{Herne2017,Herne2019}. Our single-crystal calcite rhombohedrons are grown through a precipitation process that results in somewhat uniform single crystals, with sizes ranging from three to ten microns on an edge. We noted the non-uniform rotation of the crystals when illuminated with elliptical polarization and the relationship of the rotational speeds to the crystal orientation relative to the polarization axis.

In this work we examine the equilibrium positions of calcite rhombohedrons, utilizing their asymmetry to completely predict their rotation and orientation in all degrees of freedom. We describe the torque on the crystals and show how the stable positions of zero torque relate to the linear polarization direction and beam axis. The lack of symmetry makes computational modeling time-intensive; we use an analytical approach to the progression of the electric field and propagation vector. We show experimentally the position of three-dimensionally trapped calcite and compare this with our predictions. When the crystals are trapped in $x$, $y$, and $z$ dimensions in linearly polarized light, with $z$ being along the propagation direction, the crystal will always orient itself in a particular position relative to the polarization direction and the $z$ axis. We are able to predict this orientation based on a model of the induced polarization and torque in the crystal due to the polarization and the electric field.

\section{Theoretical Background}

\subsection{Calcite Structure}

\begin{figure} [ht]
   \centering
   \includegraphics[height=5.5cm]{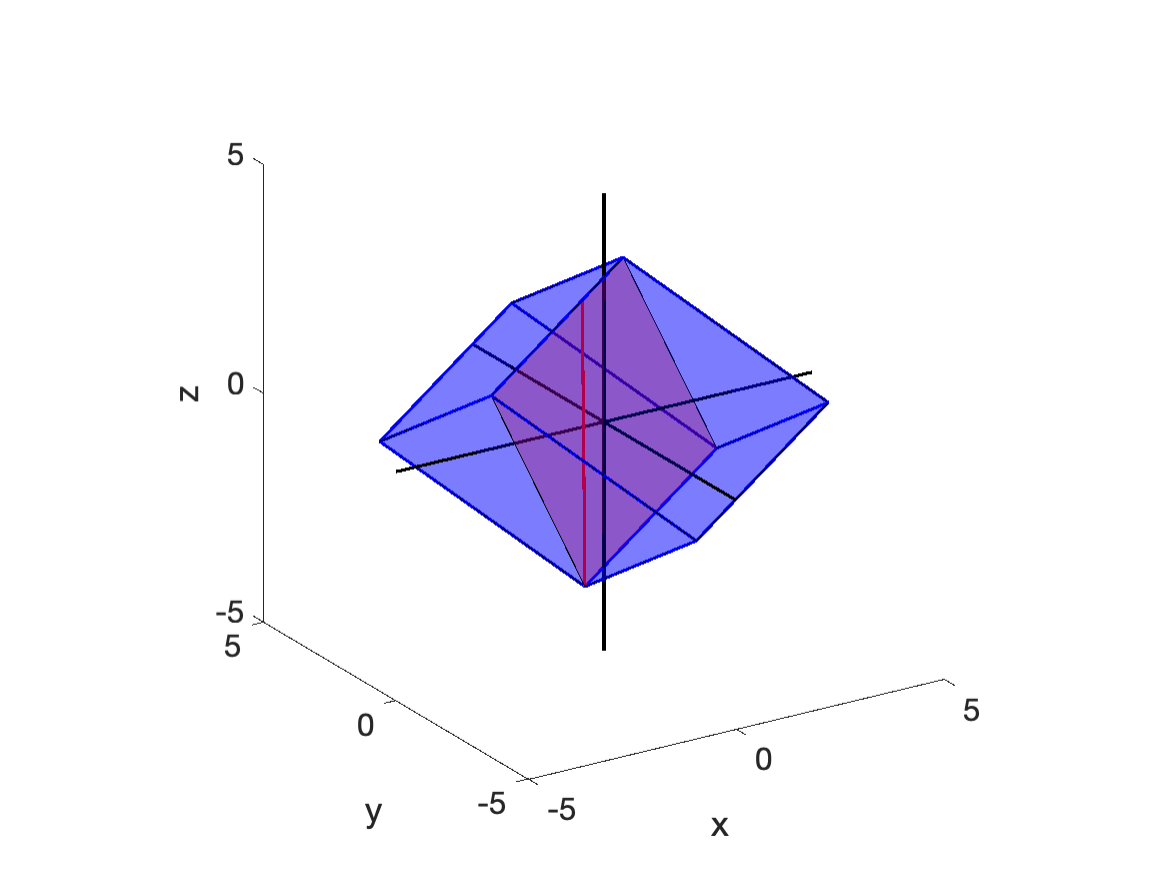}
   \caption{Calcite geometry with the blunt corners at the bottom and top and the optic axis is in red parallel to the $z$ axis. Principal section 1 (dark red) contains the optic axis (red line) and is perpendicular to two opposite faces; principal section 1 also includes the lower and upper diagonal edges.}
   \label{fig:principal1} 
\end{figure}

Calcite is one of the polymorphs of calcium carbonate, CaCo\textsubscript{3}, with negative uniaxial birefringence. It grows with a crystal structure that gives it the characteristic rhombohedral shape we see in Fig.~\ref{fig:principal1} with blunt corners at 102\degree{} and sharp corners at 78\degree. Uniaxial birefringent materials have two distinct polarization axes, referred to as the \textit{ordinary} and \textit{extraordinary} axes, and corresponding indices of refraction $n_o$ and $n_e$. For negative uniaxial crystals, $n_o>n_e$. The optic axis, defined such that a transverse electric field propagating along it will experience only one index of refraction ($n_o$), is indicated in the figure running from the lower ``blunt" corner to a point on the opposite edge ~\cite{Handbook}. There are three distinct principal sections; they contain the optic axis and are perpendicular to two opposite faces. An extraordinary axis lies in the plane of each principal section and an ordinary axis lies such that it is perpendicular to both the extraordinary direction and the principal section. Light traveling in some propagation direction $\vec{k}$ would experience the extraordinary polarization direction perpendicular to the propagation direction and in the plane of the principal section and the ordinary direction perpendicular to $\vec{k}$ and the extraordinary axis. The first principal section is the one lying in a plane containing two crystal edges. The extraordinary axis in this case will run parallel to a line from sharp corner to sharp corner, and the ordinary axis will run from blunt to blunt corner. The second and third principal section planes contain the optic axis but include only one edge. Their extraordinary axes run parallel to a line from slightly off corner to blunt corner.


\subsection{Induced Polarization}

Our theoretical approach is based on the polarizability of uniaxial crystals and the interaction of the electric field propagating through a crystal with the induced polarization. An induced polarization is given by 
\begin{equation}
\vec{P}=\epsilon_0\text{X}\vec{E}\label{pol}
\end{equation}

where $\text{X}$ is the electric susceptibility tensor, unique to each material. A fundamentally understood property is that an electrically polarized material will align its dipoles along the electric polarization of the light. When an electric field induces a polarization in a material, a torque will be generated. The torque density is  
\begin{equation}
\vec{\tau}=\vec{P}\times\vec{E}\label{torque}
\end{equation}

where $\vec{P}$ is the induced polarization in the medium and $\vec{E}$ is the electric field. We need only define the propagation direction in the crystal and the electric field vector in order to find the polarization direction and magnitude. From this we calculate the torque about all three axes and the total torque on the crystal. We then identify the crystal positions where the directional torques are a minimum.

The electric field of our incident beam generates a polarization in our calcite crystals, making it electrically polarized \cite{YarivYeh}. Due to the crystal's birefringent nature, the phase velocity of light propagating through it (and hence the index of refraction) depends on the direction of propagation and the polarization state. If we define the $z$ direction along the optic axis of the crystal, the electric field direction vector in the crystal (polarization) can be decomposed into the ordinary vector in the $y$ direction and the extraordinary vector in the $x$-$z$ plane. The resulting extraordinary polarization direction is $\hat{d}=\vec{D}/\vert\vec{D}\vert$, where the extraordinary component $\vec{D}_{induced,e}$ is shown in equation (\ref{P_e_direction_uniaxial}).

\begin{equation}
\vec{D}_{induced,e}=
\left(
\begin{matrix}
\sin(\theta)/(n_e^2(\theta)-n_o^2) \\
0 \\
\cos(\theta)/(n_e^2(\theta)-n_e^2) \\
 \end{matrix}\right) 
\label{P_e_direction_uniaxial}\end{equation}

This is a significant result, since the polarization directions for ordinary and extraordinary rays are critical for evaluating the direction and magnitude of the torque on a uniaxial crystal. The polarization as in equation (\ref{pol}) depends on the directional vector in equation (\ref{P_e_direction_uniaxial}) and on the direction of the electric field in the crystal, so the extraordinary polarization direction will be determined by (\ref{P_e_direction_uniaxial}) and the ordinary direction will be along $y$ in the index ellipsoid in the crystal. 

\begin{equation}
\vec{P}=\left(
\begin{matrix}
\epsilon_0\chi_e d_x E_x \\
\epsilon_0\chi_o d_y E_y\\
\epsilon_0\chi_e d_z E_z\\
\end{matrix}\right).
\label{polarization}\end{equation}

To understand this further, suppose light is propagating along the optic axis, so $\vec{k}$ is along $z$. Then $\theta=0$, the $x$ component of the induced polarization is zero, and the induced polarization could only be along the $z$ direction. However, there is no electric field component along the propagation direction, so the induced polarization is in fact zero! The torque is therefore also zero.

To calculate the torque we propagate the transmitted ray, electric field vector, and induced polarization through all three surfaces for the three principal sections, as the crystal is rotated by $\gamma$ and tilted by $\beta$. We recalculate the surface normals at each position. We then calculate the transmitted electric field in extraordinary and ordinary components, and determine the induced polarization. The ordinary component of the polarization lies along the ordinary axis and ordinary electric field direction for each principal section; this is the $y$ axis of the index ellipsoid. The extraordinary polarization's $z$ component is along the optic axis, and the $x$ component is perpendicular to $y$ and $z$ axes, and along the extraordinary electric field direction. We then find the vector torque density from equation (\ref{torque}) and examine the three components of torque about the $x$, $y$, and $z$ axes of the crystal, as well as the total magnitude of the torque vector. Locations of zero torque and lowest torque indicate the equilibrium positions of the crystal.

\section{Theoretical Results}
\label{sec:theoryresults}

To make our theoretical predictions we examine the torque generated by the induced polarization in the crystal.  In the trapped position, corner down, the incident rays encounter the three faces intersecting at the bottom corner. There will therefore be refraction at each of those three faces and no incident light on the upper faces. We begin our analysis with our ``zero" position such that the optic axis is along $z$ and the plane of principal section 1 includes the electric field polarization direction. We assume the crystal is at the focus of the beam and the incident rays are approximately along the beam axis. We determine the refracted angle for tilts and rotations of the crystal. The angle $\beta$ is the tilt about the $y$ axis, and the angle $\gamma$ is the rotation of the crystal about the $z$ axis. As the crystal is tilted and rotated the refracted rays for each principal section propagate in different directions. Angles for which the direction of propagation is closest to the optic axis orientation will determine the lowest torque position for each principal section.

Rays striking each bottom face of the crystal are analyzed according to the associated principal section. We set the electric field to be linearly polarized along the $x$ axis. As the crystal tilts about the $y$ axis at different $\beta$ angles, we observe a minimum torque position that is different for each principal section. In each case, when the transmitted ray is propagating closest to the direction of the optic axis, we see lowest torque. For rotation about the $z$ axis, changing $\gamma$, the lowest torque position for each principal section occurs when the extraordinary direction is parallel to the electric field polarization. 

\begin{figure} [ht]
   \centering
   \includegraphics[height=8cm]{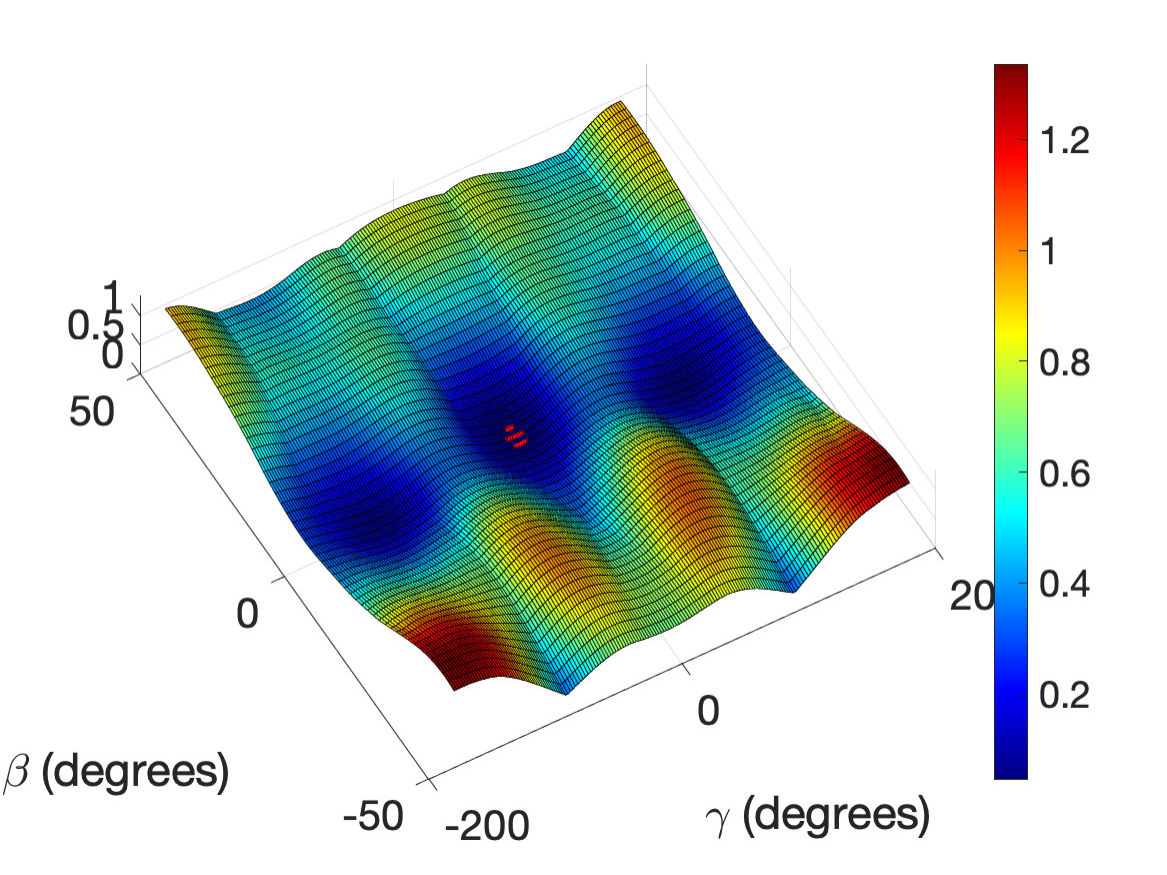}
   \caption{Total $y$ direction torque on the crystal as a function of $\beta$ and $\gamma$. The region of lowest torque is near $\gamma=0$ and $\beta=0$, indicated with off-color markers for clarity.}
   \label{fig:total_y} 
\end{figure}

\begin{figure} [ht]
   \centering
   \includegraphics[height=5cm]{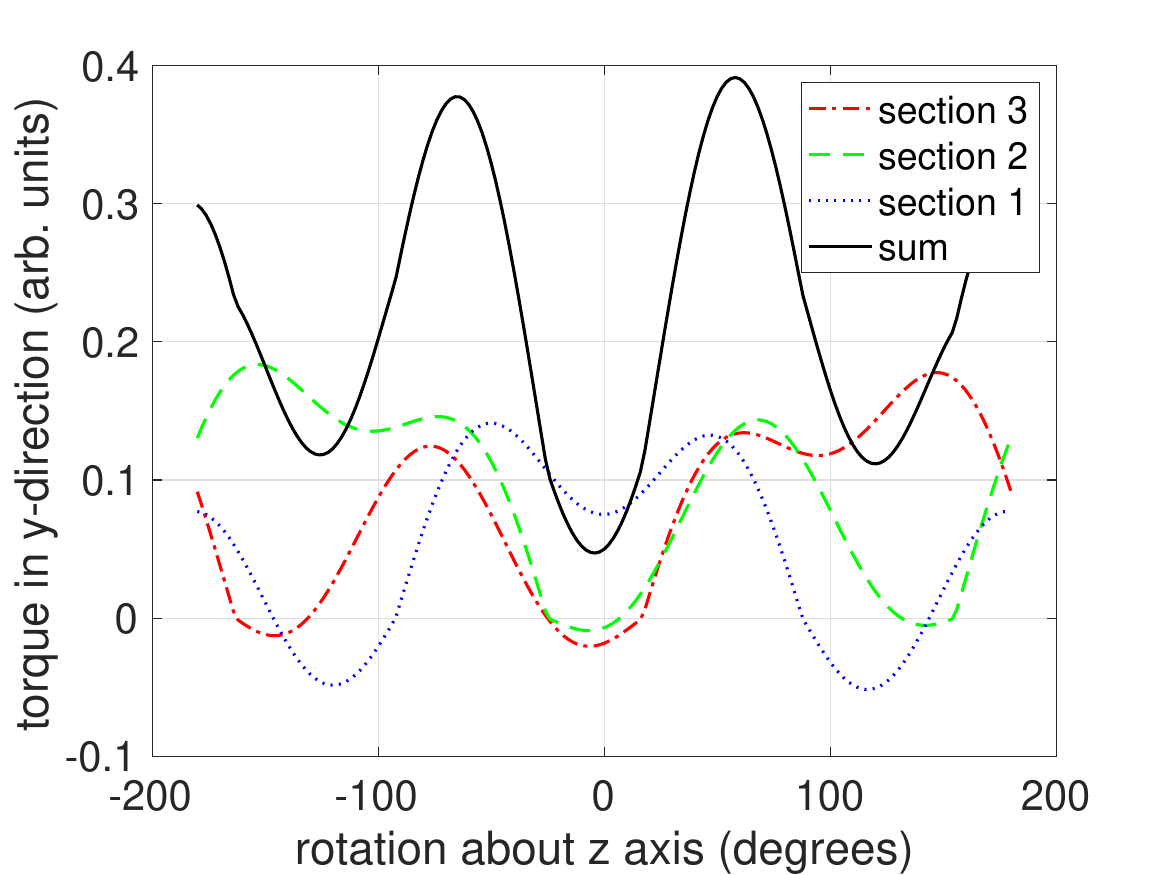}
   \caption{Torque on crystal in $y$ direction for $\beta=6\degree$ showing contributions from principal sections as crystal is rotated about $z$ by $\gamma$.}
   \label{fig:y_torque_beta_6} 
\end{figure}

The behavior of torque with tilt and rotation is shown in surface maps with $\gamma$ and $\beta$ as the horizontal and vertical axes, and the torque given by a color map. Looking first at the lowest $y$ direction torque in Fig. \ref{fig:total_y}, we see low torque values for $\beta$ and $\gamma$ near zero, and again at $\gamma = \pm130\degree$. We choose to define equilibrium for these results as when the torque is less than 4\% of its maximum value; this region occurs near $\gamma=0$ (shown with red markers) and not in the other regions. Noting again that the zero position is with the optic axis along the beam direction and the polarization along principal section 1, this is our projected result. There is a small range of $\beta$ angles with torque near zero; we discuss the implications of this in section \ref{sec:expresults}. To see each principal section's $y$-torque contribution, we show the three principal section's torques and their sum in Fig. \ref{fig:y_torque_beta_6}. The lowest sum value is at $\gamma=0$.

\begin{figure} [ht]
   \centering
   \includegraphics[height=8cm]{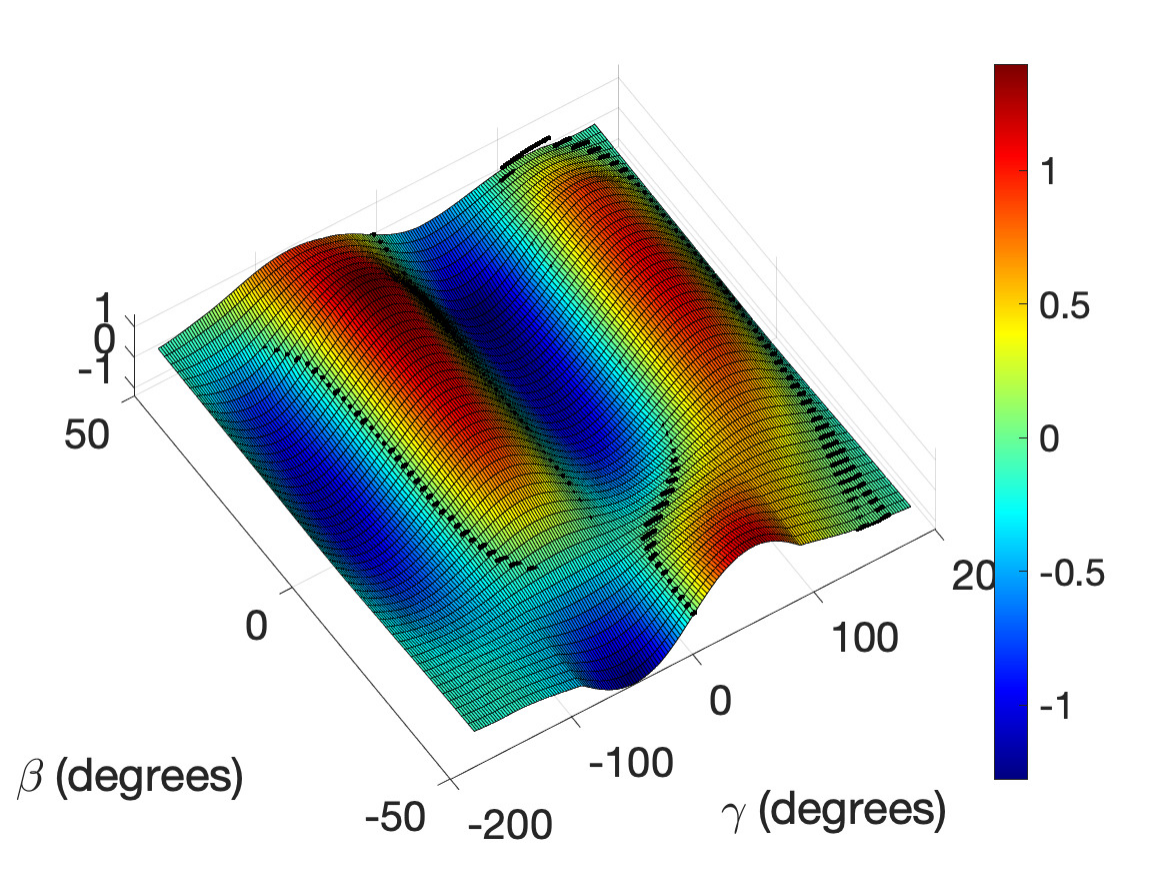}
   \caption{Total $z$ direction torque on the crystal as a function of $\beta$ and $\gamma$. The region of lowest torque is near $\gamma=0$ and $\beta=0$, indicated with off-color markers for clarity.}
   \label{fig:total_z} 
\end{figure}

\begin{figure} [ht]
   \centering
   \includegraphics[height=5cm]{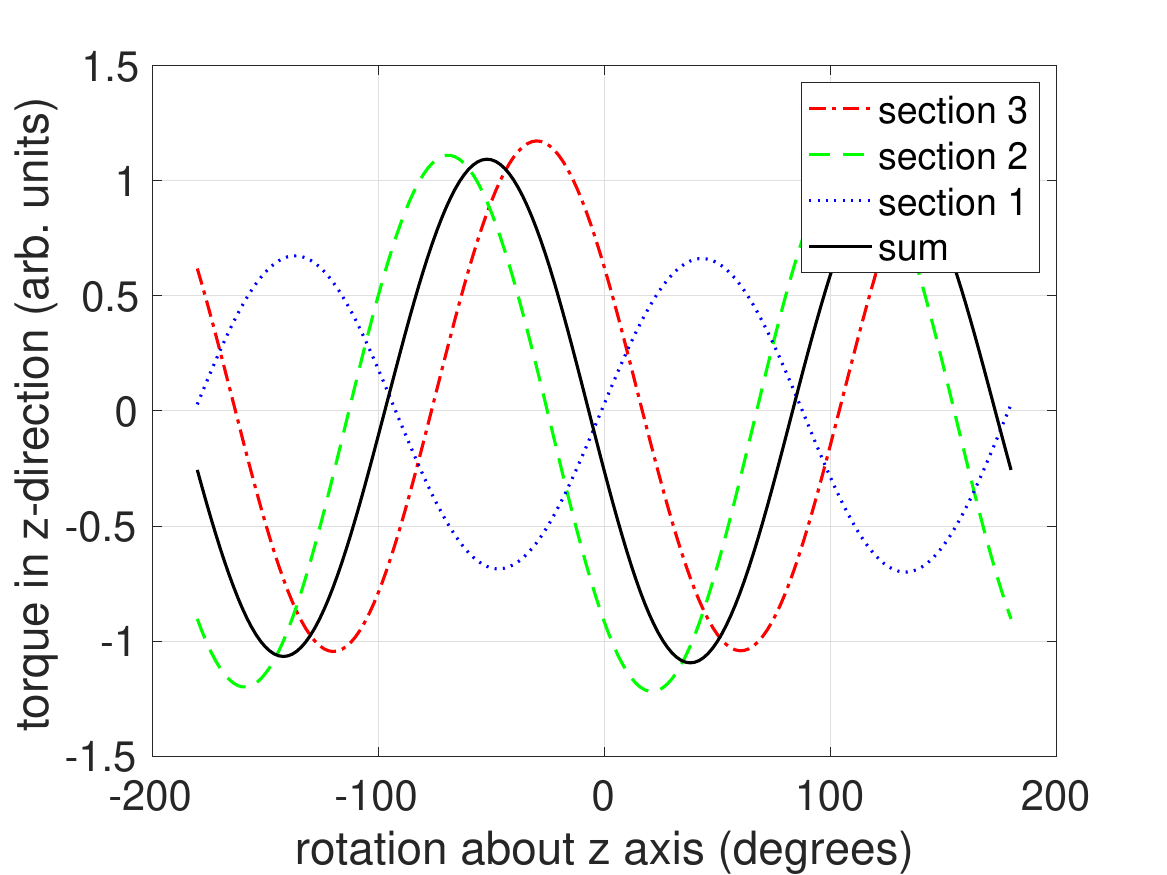}
   \caption{Torque on crystal in $z$ direction for $\beta=6\degree$ showing contributions from principal sections as crystal is rotated about $z$ by $\gamma$.}
   \label{fig:z_torque_beta_6} 
\end{figure}

\begin{figure} [ht]
   \centering
   \includegraphics[height=8cm]{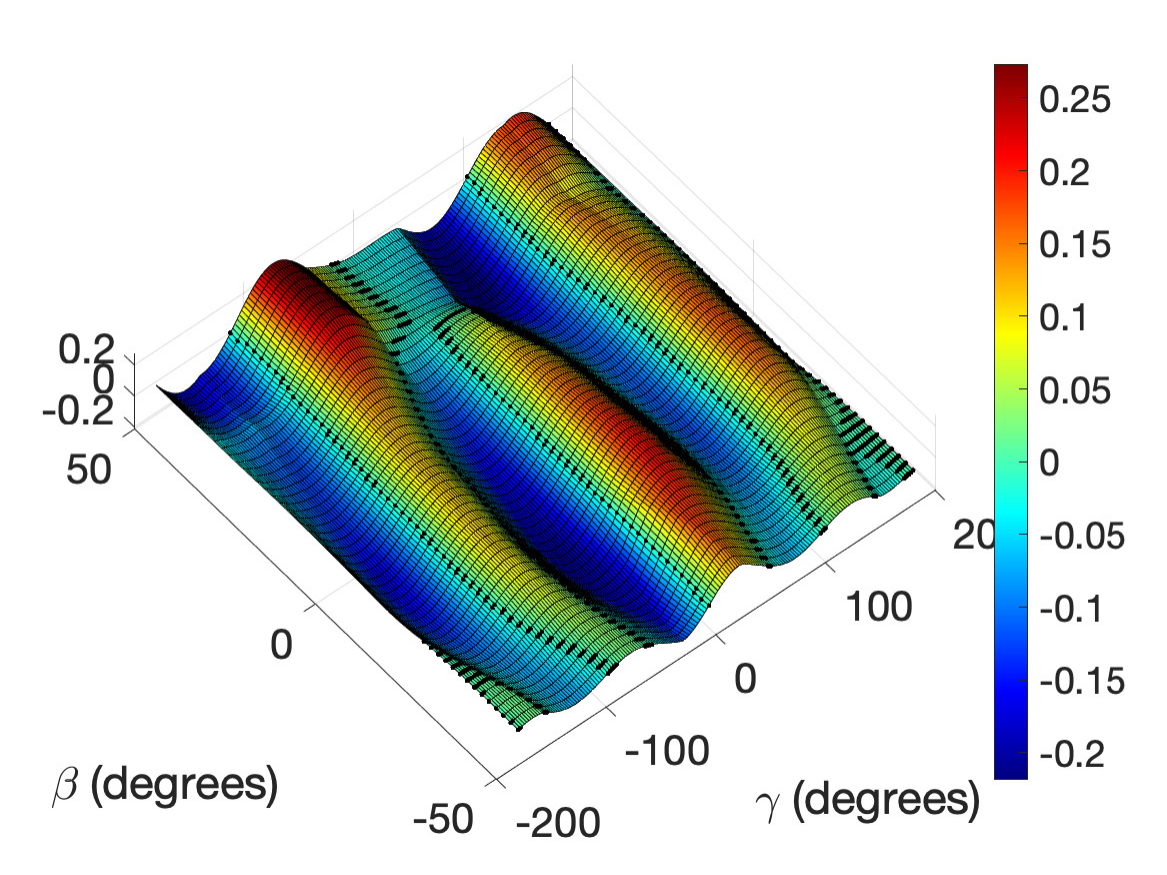}
   \caption{Total $x$ direction torque on the crystal as a function of $\beta$ and $\gamma$. The region of lowest torque is near $\gamma=0$ and $\beta=0$, indicated with off-color markers for clarity.}
   \label{fig:total_x} 
\end{figure}

Examining the same range of tilts and rotations for torque in the $z$ direction, we again see low torque at $\gamma$ and $\beta$ at zero (Fig. ~\ref{fig:total_z}) (shown with black markers). This is an equilibrium region, and here, principal section 1 is oriented with its extraordinary axis along $x$, the polarization direction. As the crystal rotates to positive (negative) $\gamma$, the torque is negative (positive), returning the crystal to equilibrium.  Further equilibrium positions can be seen near $\pm90\degree$ and $\pm180\degree$. This offers an interesting result, as it appears that there are similar equilibrium positions every 90\degree. We again examine the torque at a particular tilt in Fig. ~\ref{fig:z_torque_beta_6} and note the zeros. Looking at Fig. ~\ref{fig:y_torque_beta_6} and Fig. ~\ref{fig:z_torque_beta_6} together, the only rotation angle with equilibrium for both torque directions is $\gamma = 0$. The torque in the $x$ direction in Fig. \ref{fig:total_x} completes the picture of equilibrium, with torque equal to zero at $\gamma=0$ for all $\beta$.

Certain equilibrium positions also emerge more strongly with the crystal significantly tilted. As we model the crystal tilted so that the principal section 1 is nearly perpendicular to the beam axis, for example at 50\degree, equilibrium is at $\gamma = 0$ and $\gamma = \pm180\degree$. This won't happen experimentally while the crystal is levitated, as the $y$ torque in Fig. ~\ref{fig:total_y} shows; the crystal is unstable when tilted more than a few degrees. The crystal face is therefore normal to the incident beam only when the radiation pressure is low enough that the crystal is not levitating. In this situation the crystal face is presented to the beam such that only principal section 1 is significantly interacting with the light. When we examine the $z$ torque for section 1, there is a stable equilibrium position at $\gamma=0$ when the polarization is aligned with the extraordinary axis, as predicted in previous models \cite{Friese1998,Herne2019}.

\section{Experimental Methods}

\subsection{Crystal Growth}

We grow our crystals with a precipitate technique that results in certain crystal sizes and polymorphs depending on the molarity of solutions and time spent in precipitate. A calcium chloride solution with 0.559g of CaCl\textsubscript{2} in 100 mL of deionized water (0.05037 mol/L) is mixed slowly with 0.403g of Na\textsubscript{2}CO\textsubscript{3} in 100 mL of deionized water (0.03802 mol/L). After waiting for approximately 20 minutes, the solution is dried using a vacuum filtration system. This calcium carbonate solution can form vaterite, aragonite, or calcite. When the molar concentration, time for solution to precipitate, or temperature of solution is changed, one or another of the three polymorphs is more likely. If there is very little precipitation time, vaterite and partial crystals of calcite form. The conditions that result most often in calcite are to chill the solution and wait 20 minutes. The resulting calcite crystals are single-crystal, as can be seen in SEM pictures and in the optical tweezers imaging \cite{Herne2019}. Their sizes range from approximately 3 $\mu$m to 10 $\mu$m. Specific conditions affect the growth process, and crystals do sometimes form with unequal edge lengths. When selecting crystals to levitate, we look for those with equal edges.

\subsection{Optical Tweezers}

\begin{figure} [ht]
   \centering
   \includegraphics[height=6cm]{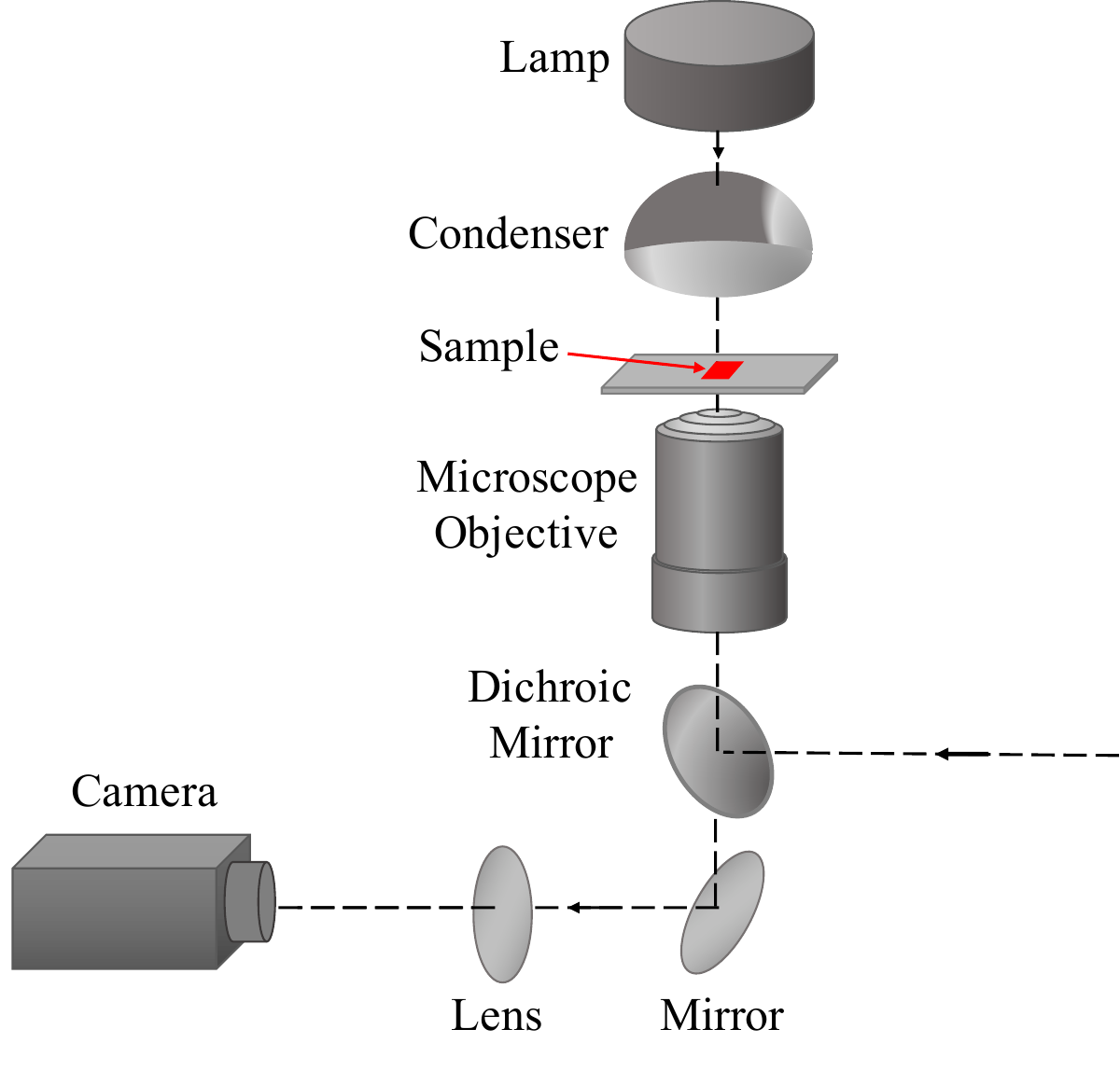}
   \caption{Experimental design of optical tweezers.}
   \label{fig:setup} 
\end{figure}

Our optical trapping arrangement (Fig. \ref{fig:setup}) uses a 100X microscope objective with a 1.25 numerical aperture. We send 660 nm light in a Gaussian mode through the objective at powers ranging from 10 mW to 80 mW. Our sample is located above the microscope objective. We image the sample onto a CMOS camera. Our sample region is bounded by 100 $\mu$m-thick double-sided tape and covered with a 100 $\mu$m-thick coverslip. The sample solution is composed of calcite crystals with deionized water and a photographic film wetting agent, which acts as a surfactant. 

\subsection{Crystal Trapping and Levitation}

With our sample in place, we search the lower surface of the sample region for single crystals (after mechanical agitation crystals still often clump together) at lower laser power. When a suitable crystal is found, it will immediately be drawn to the focus of the beam in the transverse direction. We then raise the laser power until it leaves the surface and fine-tune the power and sample position so that the crystal is trapped effectively in the middle of our sample.Trapping and levitating of larger crystals requires powers at the upper end of this range, and some crystals are too large to effectively levitate. We use a half-wave plate to set our incident polarization direction, and observe the crystal orientation at different polarizations.

\section{Experimental Results}
\label{sec:expresults}

Our single-crystal calcite has an internal structure which offers unique experimental results. Our observations and measurements consistently show the same trapped position for any crystal. We have discussed theoretically how calcite trapped and levitated in three dimensions will rest with one corner down, as in Fig.~\ref{fig:principal1}. This is dominated by the near-cubic shape and the location of the optic axis. While the most stable position for a cube levitated in optical tweezers is corner down \cite{Gauthier2000}, an object with shape birefringence should orient with its longest axis along the beam axis. In our case, however, the stable orientation is with the propagation direction along the optic axis, which positions the shorter dimension along the beam axis. As the crystal is optically trapped and levitated, it orients itself with the extraordinary axis of principal section 1 along the electric field polarization and the optic axis within a small tilt angle relative to the beam axis. As previously discussed, zero tilt is when the optic axis is aligned along the beam axis and zero rotation is with principal section 1 along the polarization direction. 

\begin{figure} [ht]
   \centering
   \includegraphics[height=4cm]{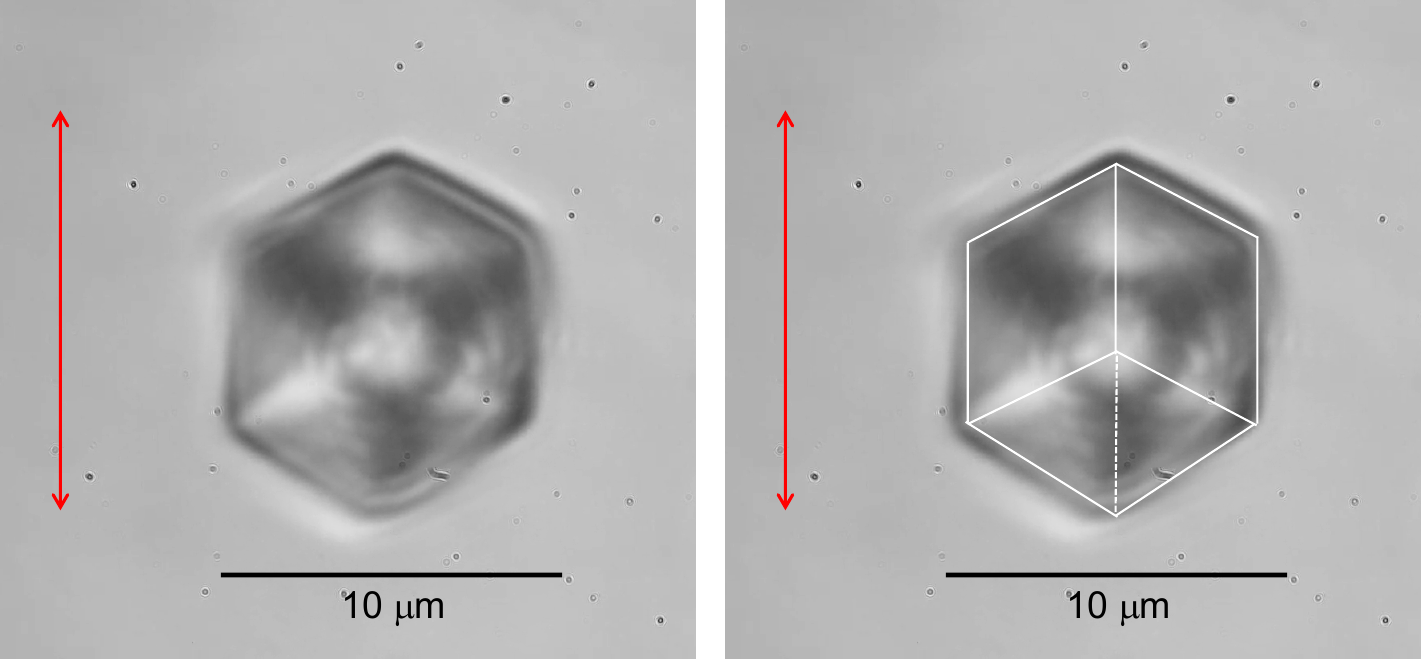}
   \caption{Calcite crystal in vertically polarized light (red). Crystal outline shown for clarity on right. Crystal is significantly larger (5 $\mu$m on edge) than depth of focus (1 $\mu$m)}
   \label{fig:calcitevertical} 
\end{figure}

A trapped calcite crystal is shown in Fig. \ref{fig:calcitevertical} with the vertical polarization direction indicated in red. Principal section 1, perpendicular to the plane of the page and including the solid and dashed line, is along the polarization direction. We are viewing directly upward along the $z$ direction and the short dimension of the crystal, which nearly aligns with the optic axis; we see the three faces and the bottom corner. The frame on the right outlines crystal edges. Tilt angle is found by measuring the lengths of the solid and dotted lines and using the geometry of the crystal to calculate an angle relative to our zero position. Equation (\ref{eqn:tilt}) is numerically solved for tilt angle $\theta$ where $L_{face}$ is the dotted line and $L_{edge}$ is the solid line.

\begin{equation}
\frac{\cos(35.5\degree+(\theta+9.1\degree))}{\cos(35.5\degree-(\theta+9.1\degree))} = \frac{L_{face}}{L_{edge}}
\label{eqn:tilt}
\end{equation}

\begin{figure} [ht]
   \centering
   \includegraphics[height=4cm]{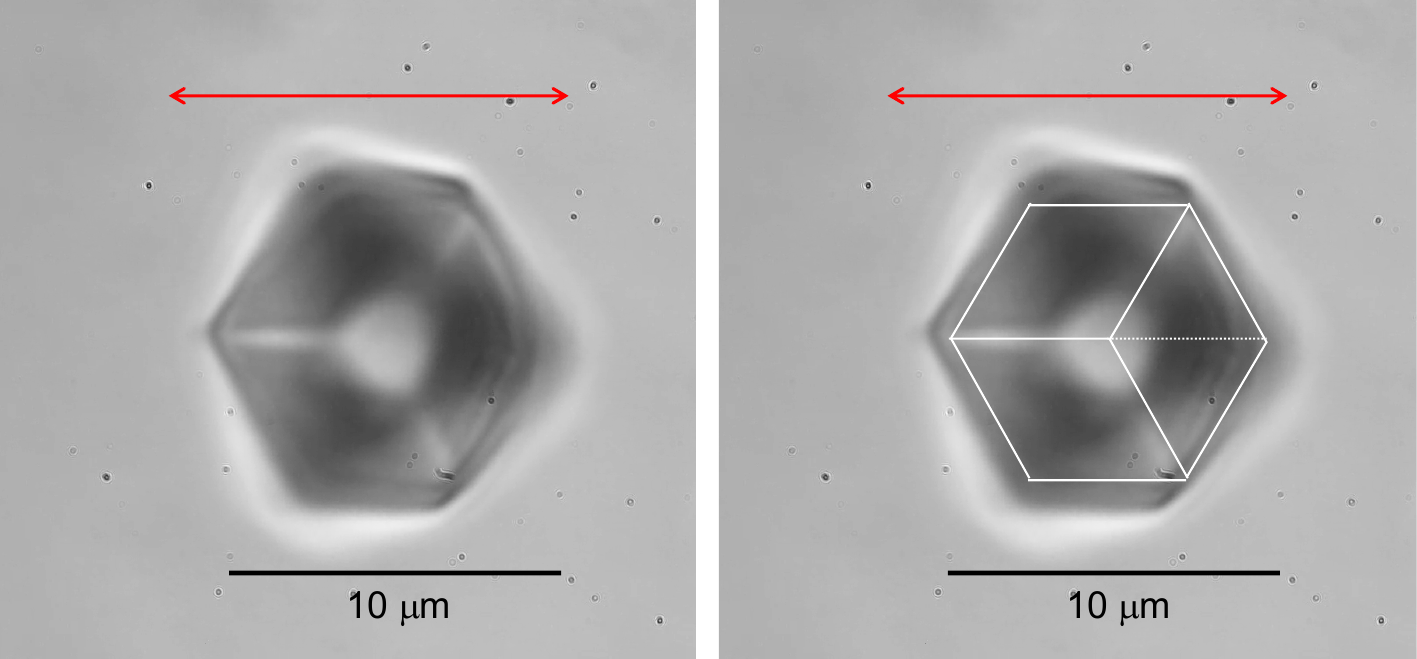}
   \caption{Calcite crystal in horizontally polarized light (red). Crystal outline shown for clarity on right.}
   \label{fig:calcitehorizontal} 
\end{figure}

 In Fig. \ref{fig:calcitevertical} the tilt is approximately $-2.0\degree\pm0.5\degree$ relative to the optic axis axis aligned with the beam axis. The same crystal with light polarized horizontally is pictured in Fig. \ref{fig:calcitehorizontal}. In each case the principal section lies along the polarization direction, as predicted.

When transitioning from resting on the bottom surface of our fluid sample region to being trapped mid-sample a crystal will tumble and rotate until its optic axis is oriented along the beam axis and one extraordinary crystal axis is oriented along the polarization direction, as discussed above. The tumbling motion is due to the $y$ and $z$ torques shown in section \ref{sec:theoryresults}. 

\begin{figure} [ht]
   \centering
   \includegraphics[height=6cm]{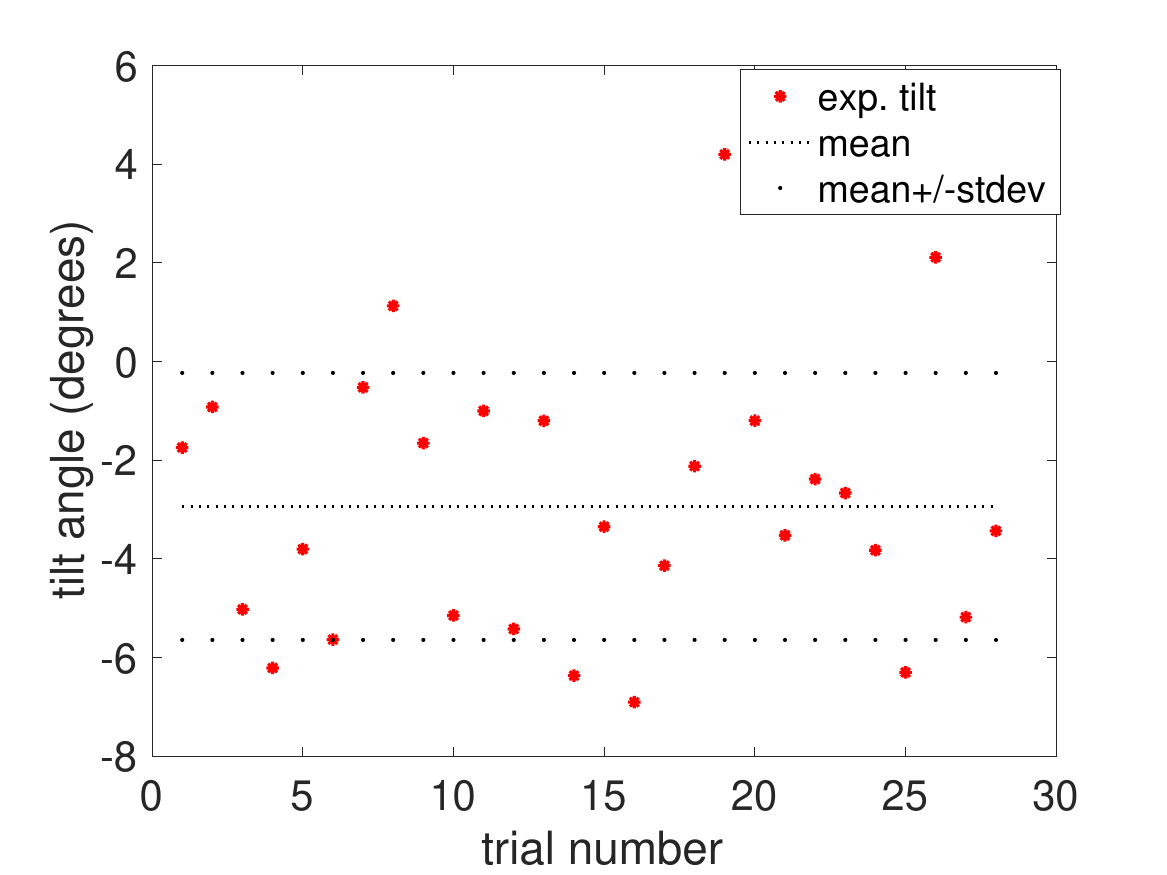}
   \caption{Tilt of calcite crystals relative to the optic axis aligned along $z$.}
   \label{fig:tiltvalues} 
\end{figure}

We also measure the tilt angle, where tilt is relative to the optic axis aligned along $z$, of several levitated crystals and show these tilts in Fig. \ref{fig:tiltvalues}. We look at the measured tilt for each of the trials shown, the mean angle, and the standard deviation of the tilt values. There is clearly a significant variation in observed tilt, from -8\degree{} to 4\degree. This can be explained by both the theoretical results and experimental conditions. As observed in Fig. \ref{fig:total_y}, equilibrium lies within a small range of tilt and rotation values; at a small rotation change, the tilt value also changes. Also, while we choose crystals that are visibly uniform on all sides, we cannot reliably confirm exact uniformity, and edge length differences would affect the resting position.

\section{Conclusions}

We have demonstrated trapping and levitating uniaxial birefringent crystals in three dimensions and three rotational degrees of freedom. The asymmetric internal and external structure of the calcite crystal extends the precision of three-dimensional trapping such that we can predict rotation about all axes. This expands previous studies on birefringent materials to those with no axes of symmetry. We use a model based on the propagation of the electric field through the crystal, induced polarization, and the resulting torque due to the interaction of laser polarization and induced polarization in the crystal. Our experimental results show equilibrium positions consistent with theoretical predictions. We note both theoretically and experimentally that there is a small range of angles for which the crystal is in equilibrium, and that the precise position is affected by the uniformity of crystal edge lengths. The crystal position can be determined to within 6\degree{} relative to the beam axis and within 5\degree{} in the transverse plane. This reflects an ability well beyond what is possible with completely symmetric birefringent spheres with 360\degree{} possible positions around all axes\cite{Stilgoe2022} or birefringent cylindrical objects with 360\degree{} rotation around one axis \cite{Singer2006}. 

Our demonstration of trapped and levitated singe-crystal calcite offers greater precision in optical positioning with the ability to predict its orientation around all rotational axes. We have shown that the position of the crystals can be determined uniquely, which offers a technique for handling these crystals as well as manipulating other objects attached to them. We are working on improved production of these crystals so their size is more accurately predicted. We also wish to explore a larger size range of crystals, from approximately $2\mu$m to $12\mu$m on an edge. We believe this work will lead to further explorations into control of complex crystals.

\section{Code, Data, and Materials Availability}

The code for this work is available at https://github.com/hernec/Calcite.

Data is available on request from the author at hernec@newpaltz.edu.
%
%


\bibliography{Herne_Calcite}   

\begin{thebibliography}{10}

\bibitem{Dienerowitz2014}
M.~Dienerowitz, L.~Cowan, G.~Gibson, {\em et~al.}, ``Optically trapped bacteria
  pairs reveal discrete motile response to control aggregation upon cell–cell
  approach,'' {\em Curr Microbiol} {\bf 69}, 669–674  (2014).

\bibitem{Nagesh2014}
B.~V. Nagesh, .~Yogesha, R.~Pratibha, {\em et~al.}, ``{Birefringence of a
  normal human red blood cell and related optomechanics in an optical trap},''
  {\em Journal of Biomedical Optics} {\bf 19}(11), 115004  (2014).

\bibitem{Pang2016}
Y.~Pang, H.~Song, and W.~Cheng, ``Using optical trap to measure the refractive
  index of a single animal virus in culture fluid with high precision,'' {\em
  Biomed. Opt. Express} {\bf 7}, 1672--1689  (2016).

\bibitem{Spudich2011}
J.~Spudich, S.~Rice, R.~Rock, {\em et~al.}, ``Optical traps to study properties
  of molecular motors,'' {\em Cold Spring Harb Protoc} {\bf 11}, 1305--1318
  (2011).

\bibitem{Mushfique2008}
H.~Mushfique, J.~Leach, H.~Yin, {\em et~al.}, ``3d mapping of microfluidic flow
  in laboratory-on-a-chip structures using optical tweezers,'' {\em Analytical
  Chemistry} {\bf 80}(11), 4237--4240  (2008).
\newblock PMID: 18442263.

\bibitem{Macdonald2003}
M.~MacDonald, G.~Spalding, and K.~Dholakia, ``Microfluidic sorting in an
  optical lattice,'' {\em Nature} {\bf 426}(12), 421--424  (2003).

\bibitem{Friese1998}
M.~E.~J. Friese, T.~A. Nieminen, N.~R. Heckenberg, {\em et~al.}, ``Optical
  alignment and spinning of laser-trapped microscopic particles,'' {\em Nature}
  {\bf 394}, 348--350  (1998).

\bibitem{Parkin2009}
S.~J. Parkin, R.~Vogel, M.~Persson, {\em et~al.}, ``Highly birefringent
  vaterite microspheres: production, characterization and applications for
  optical micromanipulation,'' {\em Opt. Exp.} {\bf 17}(24), 21944  (2009).

\bibitem{SYDG2013}
J.~L. Sanders, Y.~Yang, M.~R. Dickinson, {\em et~al.}, ``Pushing, pulling and
  twisting liquid crystal systems: exploring new directions with laser
  manipulation,'' {\em Phil. Trans. R. Soc. A} {\bf 371}, 20120265  (2013).

\bibitem{WCC2008}
K.~D. Wulff, D.~J. Cole, and R.~L. Clark, ``Controlled rotation of birefringent
  particles in an optical trap,'' {\em Appl. Opt.} {\bf 47}(34), 6428--6433
  (2008).

\bibitem{Stilgoe2022}
A.~B. Stilgoes, T.~A. Nieminen, and H.~Rubinsztein-Dunlop, ``Controlled
  transfer of transverse orbital angular momentum to optically trapped
  birefringent microparticles,'' {\em Nature Photonics} {\bf 16}, 346--351
  (2022).

\bibitem{Gauthier1997}
R.~C. Gauthier, ``Theoretical investigation of the optical trapping force and
  torque on cylindrical micro-objects,'' {\em J. Opt. Soc. Am. B} {\bf 14}(12),
  3323--3333  (1997).

\bibitem{Gauthier1999}
R.~C. Gauthier, M.~Ashman, and C.~P. Grover, ``Experimental confirmation of the
  optical-trapping properties of cylindrical objects,'' {\em Applied Optics}
  {\bf 38}(22), 4861--4869  (1999).

\bibitem{Cao2012}
Y.~Cao, A.~B. Stilgoe, L.~Chen, {\em et~al.}, ``Equilibrium orientations and
  positions of nonspherical particles in optical traps,'' {\em Opt. Exp.} {\bf
  20}(12), 12987  (2012).

\bibitem{Gauthier2000}
R.~C. Gauthier and A.~Frangioudakis, ``Theoretical investigation of the optical
  trapping properties of a micro-optic cubic glass structure,'' {\em Applied
  Optics} {\bf 39}(18), 3060--3070  (2000).

\bibitem{Higurashi1998}
E.~Higurashi, R.~Sawada, and T.~Ito, ``Optically induced angular alignment of
  birefringent micro-objects by linear polarization,'' {\em Appl. Phys. Lett.}
  {\bf 73}(21), 3034--3036  (1998).

\bibitem{Higurashi1999}
E.~Higurashi, R.~Sawada, and T.~Ito, ``Optically induced angular alignment of
  trapped birefringent micro-objects by linearly polarized light,'' {\em Phys.
  Rev. E} {\bf 59}(3), 3676--3681  (1999).

\bibitem{Singer2006}
W.~Singer, T.~A. Nieminen, U.~J. Gibson, {\em et~al.}, ``Orientation of
  optically trapped nonspherical birefringent particles,'' {\em Phys. Rev. E}
  {\bf 73}(2), 021911  (2006).

\bibitem{Herne2017}
C.~M. Herne, N.~Cartwright, and M.~Cattani, ``Determining elliptical
  polarization of light from rotation of calcite crystals,'' {\em Opt. Express}
  {\bf 25}(9), 10044  (2017).

\bibitem{Herne2019}
C.~M. Herne, J.~R. Levey, and T.~R. McCausland, ``Polarization-dependent
  non-uniform rotation rates of rhombohedral calcite,'' {\em Opt. Express} {\bf
  27}(13), 18445--18455  (2019).

\bibitem{Handbook}
M.~Bass, V.~Mahajan, E.~V. Stryland, {\em et~al.}, {\em Handbook of Optics},
  McGraw-Hill, Inc  (1995).

\bibitem{YarivYeh}
A.~Yariv and P.~Yeh, {\em Optical Waves in Crystals}, John Wiley and Sons, Inc.
   (2003).

\end{thebibliography}
\bibliographystyle{spiejour}   


\vspace{2ex}\noindent\textbf{Catherine Herne} is an Associate Professor at the State University of New York at New Paltz. She received her BS in physics and mathematics from Bryn Mawr College  in 1995 and PhD in Applied Physics from the University of Michigan in 2008.  She is interested in the pedagogy of physics teaching and in researching complex light, optical tweezers, and biomedical applications of optical tweezers. She is a member of SPIE.

\vspace{2ex}\noindent\textbf{Elaina Wahmann} is pursuing a Bachelor's degree in physics and astronomy at SUNY New Paltz. They have been actively involved in research at New Paltz for one and a half years. They have a great interest in research and plan to pursue a career in the physics field.
\vspace{1ex}


\end{document}